\documentclass[conference,10pt]{IEEEtran}
\IEEEoverridecommandlockouts
\usepackage{booktabs} 
\usepackage{url} 
\usepackage{graphicx} 
\usepackage{amsmath,amssymb,amsfonts} 
\usepackage{xcolor} 
\usepackage[font=small,labelfont=bf]{caption} 
\usepackage{textcomp} 

\usepackage{dblfloatfix}
\usepackage{cite}
\usepackage{soul} 
\usepackage{enumitem} 
\usepackage{lipsum} 
\usepackage[hang,flushmargin]{footmisc}
\usepackage{etoolbox}
\usepackage{float}
\usepackage{multirow}
\usepackage{hhline}
\usepackage{multicol}
\usepackage{amssymb}
\newlist{inlinelist}{enumerate*}{1}
\setlist[inlinelist]{label=(\arabic*)}

\usepackage{algorithmicx}
\usepackage{algorithm}
\usepackage{algpseudocode}
\usepackage{wrapfig}

\algnewcommand\algorithmicinput{\textbf{Input:}}
\algnewcommand\Input{\item[\algorithmicinput]}

\algnewcommand\algorithmicconst{\textbf{Constraints:}}
\algnewcommand\Const{\item[\algorithmicconst]}

\algnewcommand\algorithmicoutput{\textbf{Output:}}
\algnewcommand\Output{\item[\algorithmicoutput]}

\algnewcommand{\algorithmicgoto}{\textbf{go to}}%
\algnewcommand{\Goto}[1]{\algorithmicgoto~\ref{#1}}%

\algrenewcommand\algorithmicindent{0.5em}

\renewcommand{\thefigure}{\arabic{figure}}

\usepackage{tikz}
\newcommand*\circled[1]{\tikz[baseline=(char.base)]{
		\node[shape=circle,draw,inner sep=1pt] (char) {#1};}}

\usepackage{array}
\newcolumntype{L}[1]{>{\raggedright\let\newline\\\arraybackslash\hspace{0pt}}m{#1}}
\newcolumntype{C}[1]{>{\centering\let\newline\\\arraybackslash\hspace{0pt}}m{#1}}
\newcolumntype{R}[1]{>{\raggedleft\let\newline\\\arraybackslash\hspace{0pt}}m{#1}}
\newcolumntype{M}[1]{>{\centering\arraybackslash}m{#1}}
\newcolumntype{O}[1]{>{\raggedleft\arraybackslash}m{#1}}
\usepackage{makecell}
\usepackage{diagbox}

\usepackage[hidelinks, bookmarks=false]{hyperref}

\def\BibTeX{{\rm B\kern-.05em{\sc i\kern-.025em b}\kern-.08em
    T\kern-.1667em\lower.7ex\hbox{E}\kern-.125emX}}
\begin{document}

\title{ATLAS: An IoT Architecture and Secure Open-source Networking Stack for Anonymous Localization and Tracking Using\\ Smartphones and Bluetooth Beacons}

\author{
    \IEEEauthorblockN{
    Bharath Srinivas Prabakaran\IEEEauthorrefmark{1}\textsuperscript{,}\IEEEauthorrefmark{3}\thanks{\IEEEauthorrefmark{3}~These authors have contributed to this work equally.}, Felix Fasching\IEEEauthorrefmark{1}\textsuperscript{,}\IEEEauthorrefmark{3}, Juri Schreib\IEEEauthorrefmark{1}\textsuperscript{,}\IEEEauthorrefmark{3}, Andreas Steininger\IEEEauthorrefmark{1}, Muhammad Shafique\IEEEauthorrefmark{2}}
    \IEEEauthorblockA{\IEEEauthorrefmark{1}Institute of Computer Engineering, Technische Universit{\"a}t Wien (TU Wien), Austria
    \\\{bharath.prabakaran, andreas.steininger\}@tuwien.ac.at, \{e11712208, e11809908\}@student.tuwien.ac.at}
    \IEEEauthorblockA{\IEEEauthorrefmark{2}eBrain Lab, Division of Engineering, New York University Abu Dhabi, UAE
    \\\{muhammad.shafique\}@nyu.edu}
}


\maketitle

\begin{abstract}
Bluetooth (BT) has revolutionized close-range communication enabling smart capabilities in everyday devices through wireless technology. 
One of the most important sub-domains of Internet-of-Things (IoT) specializes in the usage of BT technologies to develop smart homes and environments, which include hospitals, buildings, shopping facilities, etc. to offer a wide-range of features, like instantaneous and remote access to ventilation, lighting, security, localization, and tracking. 
However, the deployment of such features in smart infrastructures are typically unaccompanied by appropriate security measures that safeguard the data and protect its users.
Towards this, we propose the ATLAS framework, which is composed of our novel IoT architecture and secure networking stack that can be used to anonymously localize and track smartphones and wearables by deploying multiple Bluetooth Low Energy (BLE) beacons across the environment. 
The proposed networking stack enables varying levels of encryption across all layers of the communication stack to ensure an easy-to-adopt, secure-by-design network architecture.
We also deploy a novel data transformation and fingerprinting-based localization algorithm, which is highly effective in localizing user devices within a given area.
The ATLAS framework is open-sourced at \textcolor{blue}{\url{https://atlas-tuw.sourceforge.io}} to enable wide-spread adoption and further research and development.
\end{abstract}

\begin{IEEEkeywords}
IoT, Smart, Environment, Security, Privacy, Anonymous, Tracing, BLE, Localization, Tracking, Bluetooth
\end{IEEEkeywords}

\bstctlcite{IEEEexample:BSTcontrol}

\section{Introduction}
\label{sec:Intro}

The next generation of automation-based quality of life improvements are expected primarily due to the advancements in the domain of Internet-of-Things (IoT)~\cite{atzori2010internet}, by enables communication amongst physical devices.
This communication can be established via various wireless platforms, like Zigbee~\cite{ergen2004zigbee}, Wi-Fi~\cite{hiertz2010ieee}, and Bluetooth  (BT)~\cite{bisdikian2001overview}, through which the user can also interact, on-demand, with devices in their the surrounding environment.
Bluetooth has proven to be quite effective as a close-range communication system, especially with the advent of Bluetooth Low Energy (BLE)~\cite{heydon2012bluetooth}, which has been developed specifically for IoT platforms and use-cases, such as smart homes, building, and environments.


\begin{figure}[t]
    \centering
    \captionsetup{singlelinecheck=false}
    \includegraphics[width = \linewidth]{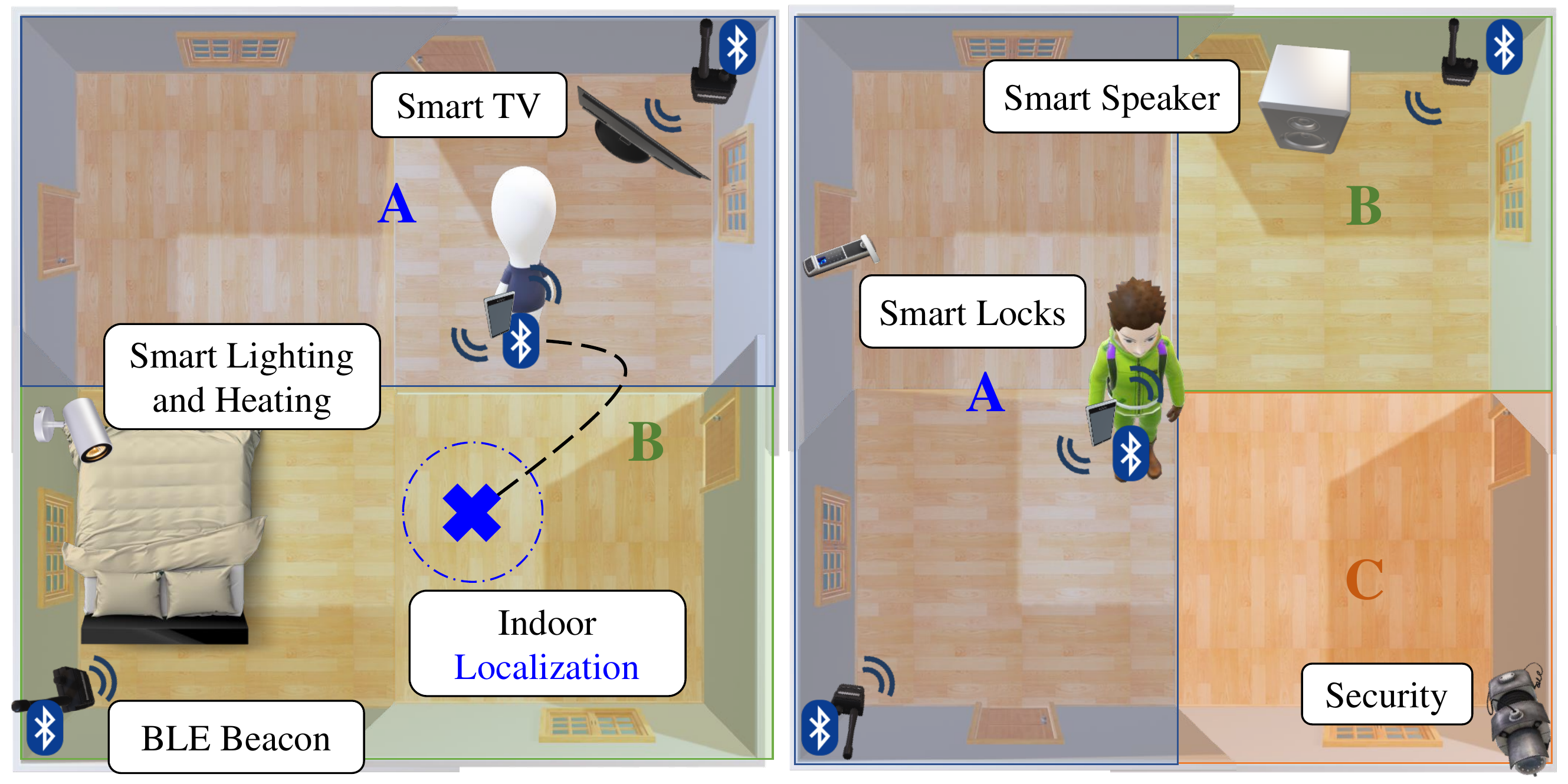}
    \caption{\textbf{An illustration of users interacting with the devices in smart environments.}{ \normalfont Connecting everyday things to the Internet can enable on-demand access to these devices via a single interface. Localization can further enhance the experience to automatically activate such devices and features when the user is in close proximity.}}
    \label{fig:UCD}
\end{figure}

Smart environments 
tend to offer a wide-range of functionalities and services, which can be controlled or operated using a single user device.
These services include remote control of heating, ventilation, and air conditioning (HVAC) systems, lighting, security systems, robots, transportation systems, smart grids, charging stations, indoor positioning, localization, tracking, etc.
An overview of such a smart environment is illustrated in fig.~\ref{fig:UCD}.
For instance, the smart television and sound system can be remotely enabled and controlled by the user's smartphone.
Similarly, connecting video cameras and other security systems to the internet enables the user to monitor their home and/or their pet during their vacation and/or work, using just a smartphone.
Furthermore, indoor localization techniques can be highly effective in automating features like heating, lighting, and ventilation, based on the user's proximity and preferred temperature settings.

Facilitating such localization techniques for larger environments, such as public buildings and shopping malls, can be quite beneficial in reducing their energy consumption and aid in emergency situations by tracking and tracing the users, e.g., contact tracing.
However, realizing such systems requires the research and development of an IoT architecture with a prime focus on the security, privacy, and anonymity of its users.
Enabling the development of a secure network infrastructure that can anonymously localize and track the user devices requires us to address the following \textbf{\textit{research~challenges}:}

\begin{enumerate}[leftmargin=*,label=(\arabic*)]
    \item How can Bluetooth low energy beacons be used to localize and track users in smart environments?
    \item What kind of IoT architecture would ensure maximum coverage area with minimum resource overhead?
    \item How do we design, develop, and deploy an easy-to-adopt secure networking stack to ensure user privacy?
\end{enumerate}

To address these challenges, we propose the \textit{ATLAS} framework with the following \textbf{\textit{novel contributions}:}

\begin{itemize}[leftmargin=*]
    \item We investigate the capability of BT beacons that can be deployed as part of our novel IoT architecture. These beacons interact with BT-enabled devices that can be used to locate and track its users anonymously.
    \item A secure-by-design architecture that prevents user-to-user communication, thereby enabling a higher level of security and privacy-preserving contact tracing capabilities.
    \item A fully open-source easy-to-adopt networking stack with varying levels of encryption at each communication interface to ensure the privacy and security of its users.
    \item A novel data transformation technique and localization algorithm that can be used to anonymously determine the location of the user and track them through the environment.
    \item Our framework is capable of localizing all user devices in range, at room-level precision, while incorporating security measures that introduce minimal communication overheads.
\end{itemize}
\section{Related Work}
\label{sec:RW}

Plenty of works have focused on device localization using various communication technologies and indoor positioning techniques, based on metrics like angle-of-arrival, time-of-flight, and signal strength.
Bahl \textit{et al.}~\cite{bahl2000radar} proposed a radio-frequency based system that relies on received signal strength indicators (RSSIs) to localize and track users inside buildings with the help of multiple base stations.
Lim~\textit{et~al.}~\cite{lim2005zero} proposed a localization algorithm that relies on the RSSIs 
\begin{inlinelist}
\item between the WiFi access points and
\item the WiFi access points and the user
\end{inlinelist}
to estimate the user location.
Diaz~\textit{et~al.}~\cite{diaz2010bluepass} proposed a signal coverage density method that creates a virtual matrix of the environment, which are composed of smaller blocks called cells. The proposed approach uses multilateration to determine the most-likely cell location of the user.
Shirehjini \textit{et al.}~\cite{shirehjini2012rfid} proposed a low-range passive Radio Frequency Identification-based (RFID) indoor positioning system that consists of carpets and peripherals embedded with RFID tags for effective data interpretation and device localization.
Kotaru \textit{et al.}~\cite{kotaru2015spotfi} presented an accurate indoor localization system that incorporates super-resolution algorithms, which accurately calculate the angle-of-arrival to determine the location of the WiFi users to decimeter-level accuracy.
Ashraf \textit{et al.}~\cite{ashraf2020minloc} proposed the use of several convolutional neural networks to analyze geomagnetic field patterns and determine the exact user location.

\begin{figure*}[t]
    \centering
    \captionsetup{singlelinecheck=false}
    \includegraphics[width = \linewidth]{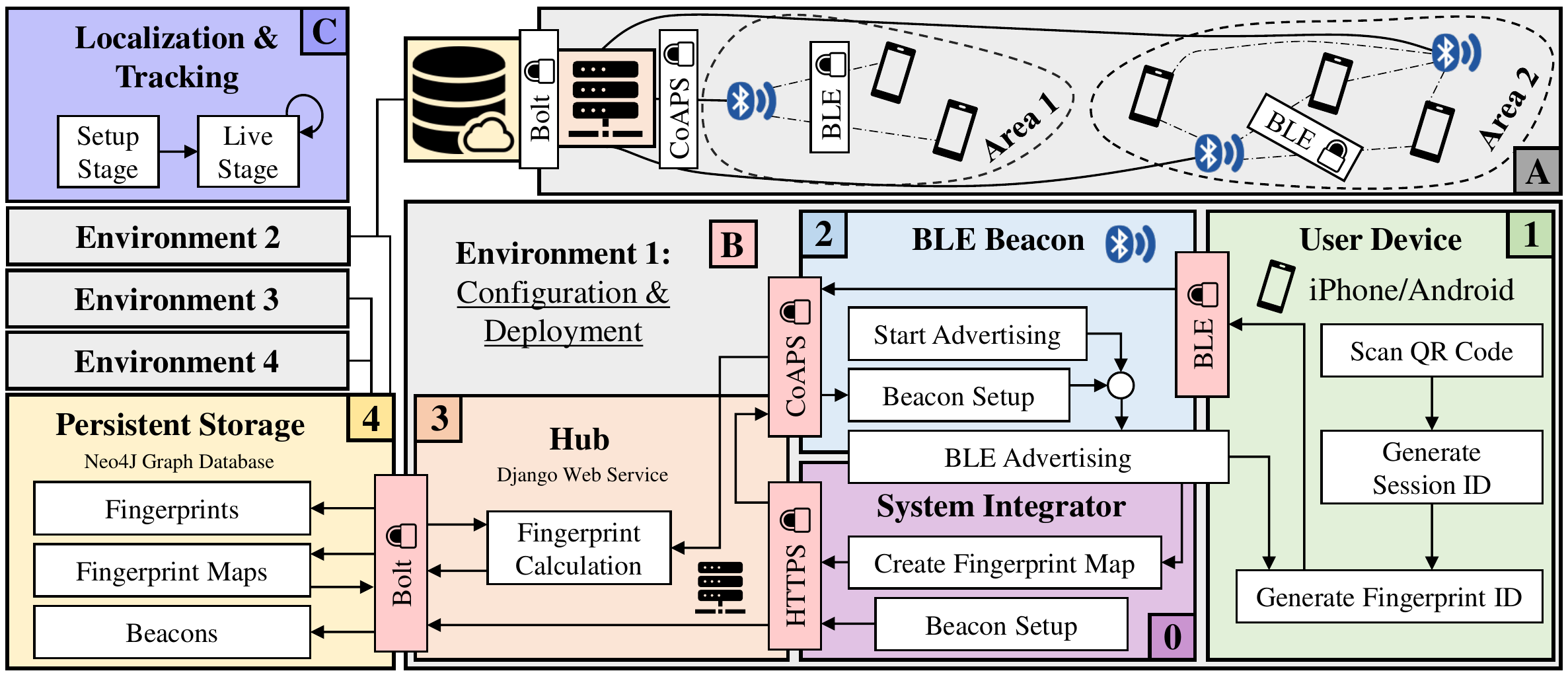}
    \caption{\textbf{An overview of the proposed ATLAS framework.}{ \normalfont Note, the novel IoT architecture depicted on top illustrates the lack of communication among different user nodes, which provides a higher layer of security against malicious devices, which interface solely with the beacons.
    After the initial setup stage, the user device interacts with the BLE beacons, periodically, which subsequently transmit data to the hub that is processed to generate location information that can be used to track them. Each layer of the networking stack is encrypted and secured to ensure the privacy, security, and anonymity of its users.}}
    \label{fig:ATLAS}
\end{figure*}

However, more recently, BT-based localization has proven to be quite effective, especially with the advent of BLE.
For instance, Zafari \textit{et al.}~\cite{zafari2017ibeacon} proposed a BT-based indoor localization system that relies on novel filtering and averaging approaches, which are used to reduce the fluctuations in RSSIs to improve the localization accuracy.
Martin \textit{et al.}~\cite{martin2014ibeacon} also presented a BLE-based localization technique around the iBeacon technology protocol invented by Apple. 
This technology is currently in use in the Apple AirTag~\cite{Airtag} devices, which can be leveraged by its users to locate and track things with the help of BT and the network of apple smart phones around the world.
Google has had similar technology ventures in indoor localization via the Eddystone project~\cite{Geddy}.
However, these devices act as passive beacons without any trust verification for the beacons, require internet access, and user device identification.
To ensure the privacy of the users and the environmental BLE beacons, Fawaz \textit{et al.}~\cite{fawaz2016protecting} proposed a device-agnostic system that can enable the users and system administrators to control who can discover, scan, and connect to their devices.
On the other hand, Kolias \textit{et al.}~\cite{kolias2017breaking} analyzed a wide plethora of BLE nodes to identify vulnerabilities can track the users, their behavior, and spoof the network.

These types of techniques primarily focus on a single aspect, which is either localization and indoor positioning or security and privacy.
To the best of our knowledge, our ATLAS framework is the first to combine both aspects to ensure that the privacy and security of its users are not compromised while ensuring that the user devices are localized and tracked periodically, using our localization algorithm. 
Note, the use of other localization algorithms is orthogonal to the use of the proposed localization technique.
Similarly, the use of other secure networking stack protocols can be easily incorporated into the ATLAS framework based on the system requirements and specifications. 

\section{Our ATLAS Framework}
\label{sec:ATLAS}

Fig.~\ref{fig:ATLAS} illustrates an overview of our ATLAS framework including the proposed novel IoT architecture and the secure, open-source, and easy-to-adopt networking stack, which can be used to anonymously track users using our novel localization algorithm.
During the initial setup stage, network administrators and system designers are required to deploy the BLE beacons, interface them with the hub, and generate a set of reference points known as the \textit{fingerprint map}, which can be used to subsequently localize the users to specific ``areas'' of the environment.
The entire networking stack is fully encrypted using various encryption techniques to ensure the privacy, security, and anonymity of its users.

\subsection{IoT Architecture}
\label{subsec:IoT}

According to a $2018$ mobile consumer survey, smartphone penetration in developed nations and regions, such as the EU and the US, averages at $91\%$, i.e., on average, $91\%$ of the population owns a smartphone~\cite{deloitte2018}. 
Due to their wide-spread adoption and the low energy requirements of BT, we propose to use these devices to actively localize and track the users by communicating with them periodically, especially since the BT module on such devices are typically active communicating with other wearables such as smart-watches and wireless headphones.
We do this by establishing an initial communication via a quick response (QR) code (generated by the system integrator; block~0 in Fig.~\ref{fig:ATLAS}), which the users scan through the custom application to interface with the smart environment (block~1).
This one-time user interaction establishes device communication with nearby BLE beacons~\cite{maier2017comparative} that are deployed as part of the environment, as shown in Fig.~\ref{fig:ATLAS}.
The beacons advertise and communicate with the user devices, periodically, to estimate the received signal strength indicator (RSSI) for each user in the network (block~2).
The beacons subsequently transmit this data to the hub (block~3), which connects to all beacons in the ``environment'', to process and determine the live \textit{fingerprints} of the users, along with a timestamp regarding their communication, to enable tracking the user through the environment (see Section~\ref{subsec:LnT}).
The system integrator (block~0), which is active only during the setup stage, is responsible for determining the initial \textit{fingerprint map}, which serves as a reference for localizing user devices.
The data collected across multiple environments, by their respective hub nodes, are transmitted to a persistent cloud-based storage system (block~4), which is stored for a period of $28$ days as per the privacy regulations of the EU~\cite{voigt2017eu} in order to enable anonymous tracking and contact tracing capabilities.

\subsection{Secure Networking Stack}
\label{subsec:SNS}
To ensure the privacy, security, and anonymity of the users, we propose to secure each communication layer in the network stack with varying levels of encryption.
The proposed secure networking stack is open-source and available online at \textcolor{blue}{\url{https://atlas-tuw.sourceforge.io}}, which can be easily adopted for similar IoT architectures to ensure wide-spread adoption and further research in this domain.
The initial pairing of the user device with the BLE beacons takes place using out-of-band (OOB) pairing with the OOB secret hidden in the QR code, which needs to be scanned using the custom application.
The transmitted data packets are BLE link-layer encrypted, to ensure that the packet sniffers are unable to extract user information.
Next, the data transmission between the hub and the beacons is completed using the constrained application protocol (CoAP)~\cite{bormann2012coap} combined with the datagram transport layer security (DTLS)~\cite{kothmayr2013dtls} protocol to provide a higher layer of security in order to prevent eavesdropping, tampering, and message forgery.
During beacon deployment and system setup, symmetric keys are pre-shared among the beacons and the hub, to establish a secure line of DTLS-PSK communication, followed by which the public key infrastructure (DTLS-PKI) is available to provide reliable beacon authentication to the hub~\cite{seoane2020performance}.
The hub interfaces with the cloud-based storage device using the Bolt protocol~\cite{Bolt} with a TLS encryption to ensure the security requirements of our system.
Similarly, the database stored in the cloud is also encrypted.
Investigating the security aspects and vulnerabilities of the implemented communication and encryption protocols lie beyond the scope of this work.


\subsection{Localization and Tracking}
\label{subsec:LnT}

\begin{figure}[t]
    \centering
    \captionsetup{singlelinecheck=false}
    \includegraphics[width = \linewidth]{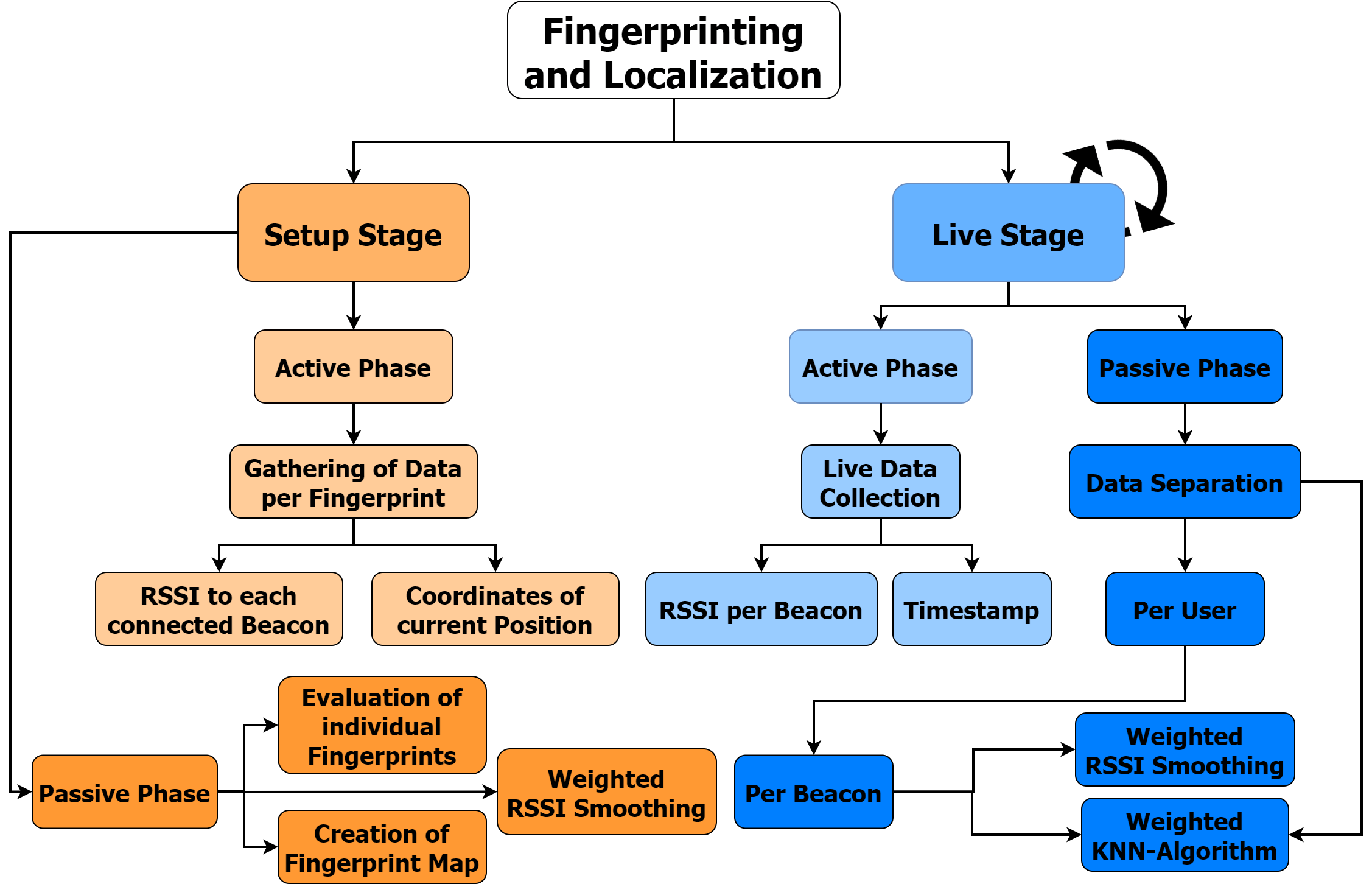}
    \caption{\textbf{An overview of the localization technique used in the ATLAS framework.} The proposed technique works in two stages; the setup stage, during which an initial fingerprint map is calculated to aid in user device localization, and the live stage, which continuously executes at run-time to actively locate and track users.\label{fig:LnT}}
\end{figure}

We propose to use the \textit{fingerprinting} technique, shown in Fig.~\ref{fig:LnT}, for localizing user devices, which works in two stages:

\subsubsection{Setup stage}
\begin{figure}[h]
    \centering
    \captionsetup{singlelinecheck=false}
    \includegraphics[width = \linewidth]{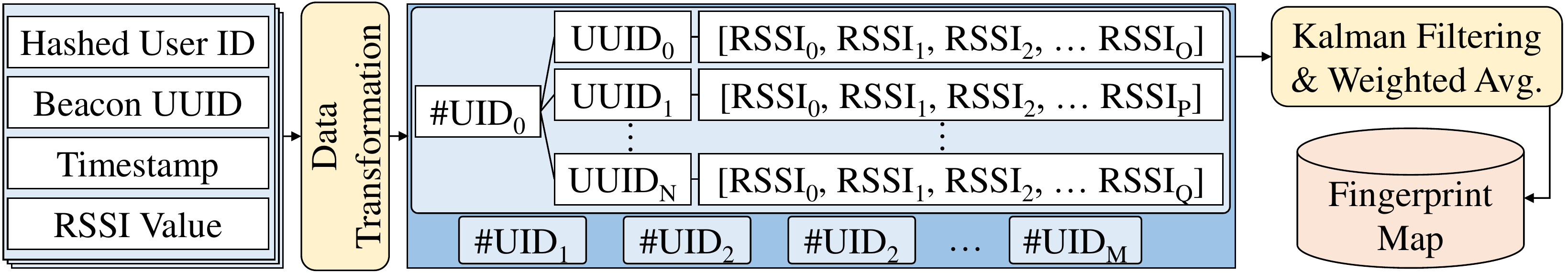}
    \caption{\textbf{An overview of the data transformation technique used to generate fingerprints and the fingerprint map.}{ \normalfont After collating the data in the required format, the data is filtered and processed using weighted averaging to generate the fingerprint map.}}
    \label{fig:Transform}
\end{figure}

Our technique relies on the existence of a predefined set of reference points, which requires our system to undergo an initial setup stage composed of active and passive phases.
During the active phase, a smartphone with our customized application and administrator privileges is used to generate the set of reference points, using the RSSI values from each beacon and the device's coordinate values, at different locations in the environment (block 0 in Fig.~\ref{fig:ATLAS}).
These reference points should be, typically, distributed evenly throughout the environment in order to get optimal results; a non-evenly distributed set of references may lead to incorrect localizations.
In the passive phase, the data collected during the active phase is sorted, based on their timestamps, and filtered based on their RSSI values per beacon, using a Kalman filter to smoothen potential outliers.
The processed data act as individual \textit{fingerprints} that contain the RSSI per beacon at the reference coordinate.
These fingerprints are used to generate the \textit{fingerprint map}, which is stored as a database on the cloud.
Fig.~\ref{fig:Transform} illustrates an overview of this approach.
Note, any significant changes to the environment, i.e., changes to the existing infrastructure, which could negatively affect the localization, can be addressed by re-executing the initial setup stage to generate a new set of reference points.

\begin{algorithm}[t]
    \normalsize
    \caption{The Localization Algorithm}
    \label{Algo1}
    \begin{algorithmic}[1]
        \Input $\text{User Fingerprint Array} (FP_U)$
        \Input $\text{Fingerprint Map} (FM)$
        \Output $\text{Array of User Locations} (L_U)$
        \State $L_U = Dict(\{\});$
        \State $boolArray = [];$
        \For{$U$ \textbf{in} $FP_U$}
        \State $validFP = Dict(\{\});$
        \For{$fp$ \textbf{in} $FM$}\label{m1}
        \State $fpB_A = fp.beaconArray()$
        \State $UB_A = U.beaconArray()$
        \State $boolArray = fpB_A$ $\&$ $UB_A$
        \If{$size(boolArray) < 2$}
        \State $\textbf{continue}$
        \EndIf
        \State $validFP[fp] = fp.beaconArray()$
        \EndFor\label{m2}
        \State $wtfp = Dict(\{\});$
        \For{$vfp$ \textbf{in} $validFP$}\label{m3}
        \State $fpw = 0$
        \State $wt = 0$        
        \For{$beacon$ \textbf{in} $U.fingerprint$}
        \If{$beacon.uuid$ \textbf{in} $vfp.beaconArray()$}
        \State $val1 = beacon.rssi$
        \State $val2 = vfp.beaconArray(beacon.uuid).rssi$
        \State $wt = wt + (val1-val2)^2$
        \EndIf
        \EndFor
        \State $wt = wt/size(boolArray)$
        \State $wt = sqrt(wt)$
        \State $fpw = fpw + wt$
        \State $wtfp[vfp] = fpw$\label{m4}
        \EndFor
        \State $L_U[U] = max(wtfp)$
        \EndFor
        \State \textbf{return} $L_U$\label{m5}
    \end{algorithmic}
\end{algorithm}

\begin{figure*}[t]
    \centering
    \captionsetup{singlelinecheck=false}
    \includegraphics[width = \linewidth]{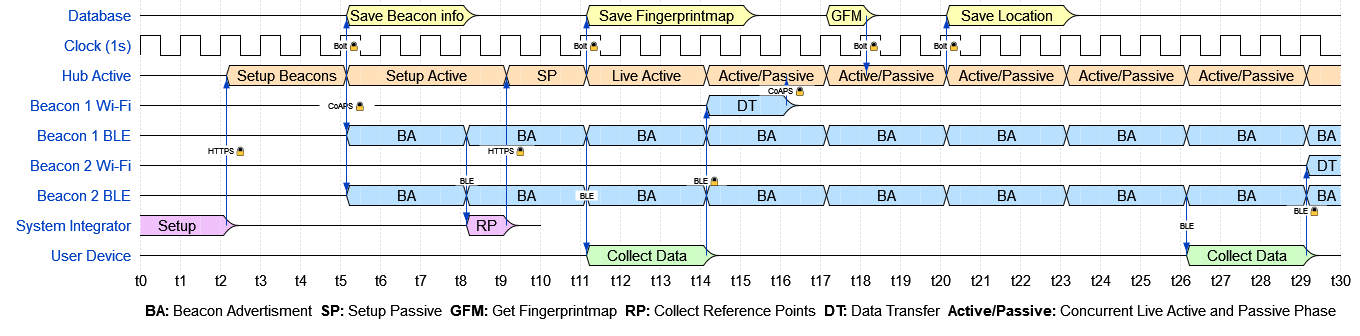}
    \caption{\textbf{Timing analysis of our ATLAS framework.}{ \normalfont Collecting the reference points depends on the environment and the number of beacons deployed; for the purpose of simplicity, we consider $1$ clock-cycle duration. The beacons connect to the user device every $15$s and since a beacon-user interaction requires up to $3$s, theoretically, a total of up to $40$ user devices can connect to a single BLE beacon.}}
    \label{fig:TD}
\end{figure*}

\subsubsection{Live stage}
After the setup stage, the live stage of our localization technique is executed continuously at run-time to enable interactions with new users and track them through the smart environment.
Like in the previous stage, data is collected during the active phase of this approach, and processed during the passive phase to perform localization.
During the active phase, the beacons periodically advertise to the connected user devices and collect the RSSI information for each device in the environment.
The data processing, on the other hand, is completed during the passive phase of the live stage, wherein the data collected during the active phase is used to construct \textit{fingerprints}, similar to the approach presented in the passive phase of the setup stage.
Next, we fetch the fingerprint map from the database to evaluate the location of the user by comparing the fingerprints obtained from the user to the fingerprints in the map.
We do this by iteratively comparing each generated user fingerprint with fingerprints in the fingerprint map. 
First, we determine the set of common beacons between the two compared fingerprints, which if less than 2 are automatically discarded, and store them in an array of valid fingerprints ($validFP$) (lines~\ref{m1}-\ref{m2}).
Next, we determine the proximity of a user to a given fingerprint by estimating the euclidean distance between the RSSI value for each beacon in the user fingerprint array and the RSSI value of the beacon from the valid fingerprint array.
These distances are accumulated per beacon to generate a weight distribution for each fingerprint in the fingerprint map, thereby determining a proximal location for each user, based on their fingerprint (lines~\ref{m3}-\ref{m4}).
After iterating through each generated fingerprint per user, an array of the proximal locations of each user, along with their hashed user ID, is returned by the algorithm (line~\ref{m5}), which is subsequently stored in the cloud, as discussed in Section~\ref{subsec:IoT}.
Algorithm~\ref{Algo1} discusses the pseudo-code for the proposed localization technique.

\section{Experimental Setup}
\label{sec:ES}

We have used the following list of equipment for the illustrating the results presented in Section~\ref{sec:Results}: $9$ ESP-$32$ beacons, which are WiFi and BLE enabled, $1$ Hub (Dell XPS $13$ with an Intel i$5-6200$U processor with $8$GB RAM and $256$GB SSD, running Ubuntu $21.04$), and a WiFi-Access Point (Huawei $4$G Router B$528$).
The beacons are deployed in various configurations across two floors of a building, an overview of which is provided in Section~\ref{sec:Results}.
We use a $2019$ iPad Pro as our system integrator (block 0 in Fig.~\ref{fig:ATLAS}) during the active phase of the setup stage, which subsequently transmits the collected data to the hub to generate the \textit{fingerprint map} during the passive stage.
The final deployed environment is tested during the live stage using $4$ smartphones and tablets, an iPhone $8$ Plus, a $2016$ iPhone SE, a $2019$ iPad Pro, and a $2019$ iPad (all of them running iOS). 
The mobile application is programmed in Swift utilizing the ARKit, a software development kit (SDK) that associates beacon RSSI values with user device coordinates, and the Bluetooth core, which is essential for communicating with our BLE beacons.
Similarly, for android platforms, Google offers an SDK called ARCore, which can be used to design similar applications.
The firmware for the ESP-$32$ BLE beacons are compiled in C with the \textit{NimBLE} and \textit{LibCoAP} libraries for Bluetooth and WiFi communication.
Python is used as the default programming language on the hub to realize the Django web service, the CoAPS, HTTPS, and Bolt communication protocols.
For simplicity, the persistent cloud-based storage system is realized as a Virtual Private Server (VPS) in the hub, hosted by Netcup, to store the encrypted fingerprint map database.
The VPS is allocated $1$ core, $2$GB RAM, and $20$GB of SSD space to run a Debian bullseye $11$ operating system, which executes the Java-based Neo4J graph database that is ideal for storing graph-based information, such as the timestamped fingerprint maps generated by our framework. 
The Django-neomodel Python library is used as an object graph mapper for communicating with the Neo4J database.

\section{Results and Discussion}
\label{sec:Results}

\setcounter{figure}{7}
\begin{figure*}[b]
    \centering
    \captionsetup{singlelinecheck=false}
    \includegraphics[width = \linewidth]{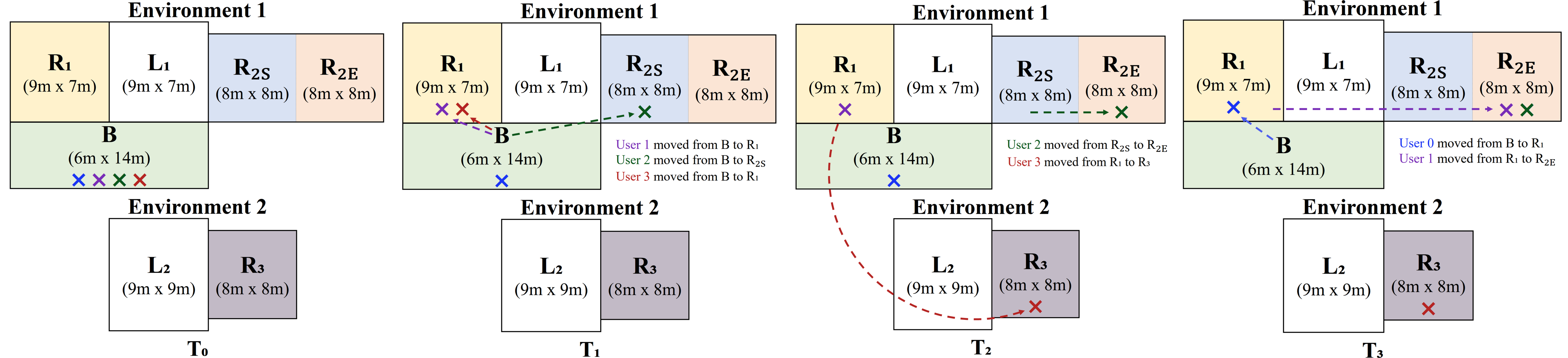}
    \caption{\textbf{Localization and tracking analysis of four users in a multi-environment scenario.}{ \normalfont Our ATLAS framework is capable of recording and processing the data of multiple users across multiple environments, which are interconnected.}}
    \label{fig:LTME}
\end{figure*}

Fig.~\ref{fig:TD} presents an overview of the end-to-end timing analysis of our ATLAS framework.
After the initial beacon configuration and deployment, the system integrator is also responsible for collecting the reference points, the duration of which heavily relies on the environment and the number of beacons deployed. For the sake of simplicity, we consider this duration to be $1$s in our analysis.
Once the beacons are setup and deployed, the hub configures them to immediately start advertising, but the users can only connect to them once the setup stage is complete as shown by the clock at $t_{11}$.
During this time, the beacon information is stored in the database and the fingerprint map is calculated, which is also transferred to the database in the next subsequent clock-cycle.
Since the round-trip communication between the user and the beacon requires $3$s, primarily due to the encryption layer, the BLE advertisements and the active/passive phases of the live stage in the hub are synchronized accordingly, as shown.
Theoretically, every beacon can connect to up to $8$ user devices in parallel using BLE and since we repeatedly ping the discovered devices every $15$s, we can locate up to $40$ devices in an area.
Similarly, we can also increase the localization frequency of the discovered devices by trading-off on the number of locatable devices in the environment.
The data collected by the beacon is transmitted to the hub for processing and comparison with the fingerprint map during the next passive phase of the hub.
This process executes continuously until the system is manually shutdown or paused by the administrator.

\setcounter{figure}{5}
\begin{figure}[t]
    \centering
    \captionsetup{singlelinecheck=false}
    \includegraphics[width = \linewidth]{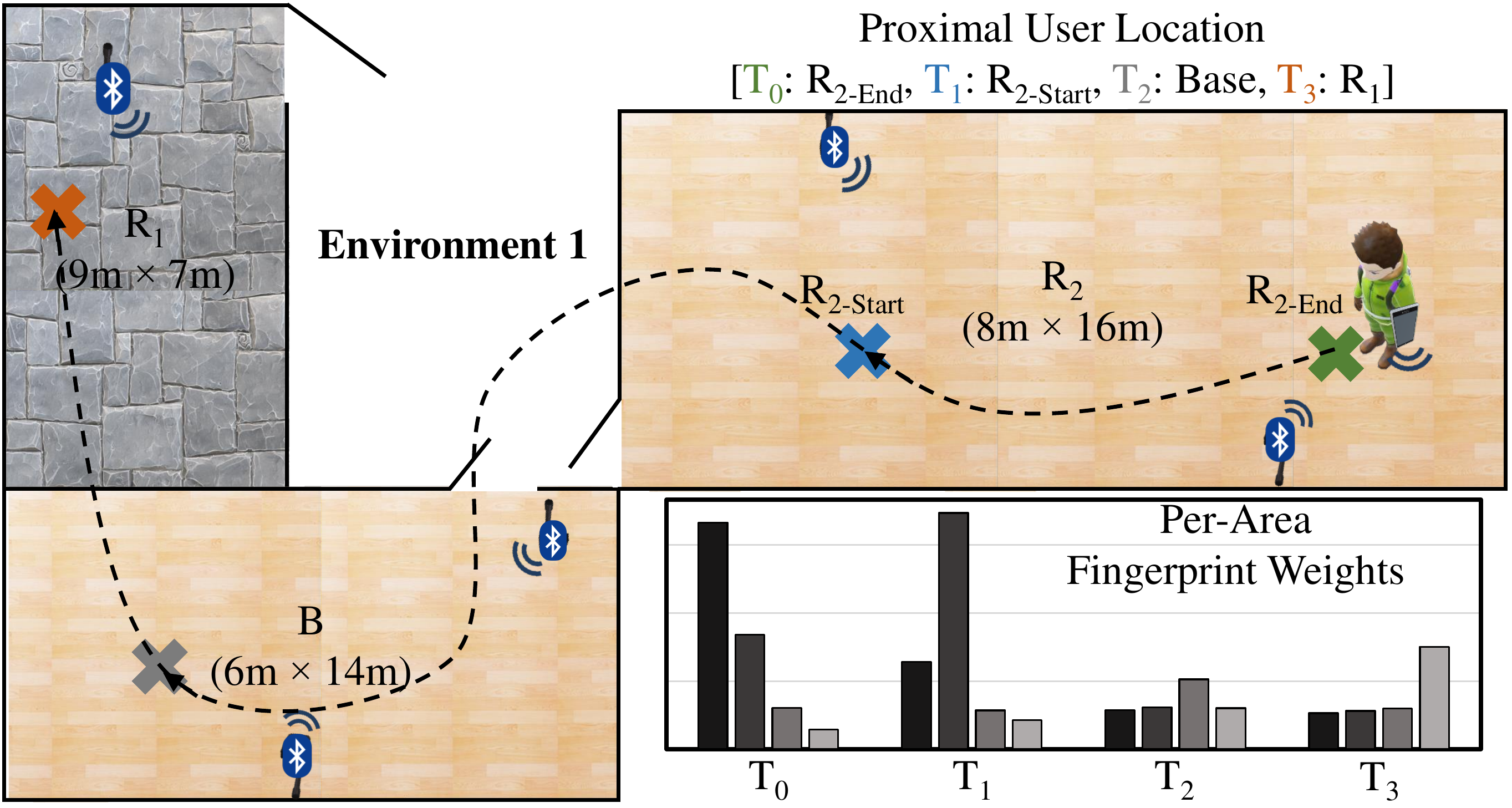}
    \caption{\textbf{The floor-plan and the beacons deployed in Environment~1.}{ \normalfont With the user generated data for fingerprints and their timestamps, we can localize and track their devices across our environment.}}
    \label{fig:FP}
\end{figure}

\setcounter{figure}{6}
\begin{figure}[t]
    \centering
    \captionsetup{singlelinecheck=false}
    \includegraphics[width = \linewidth]{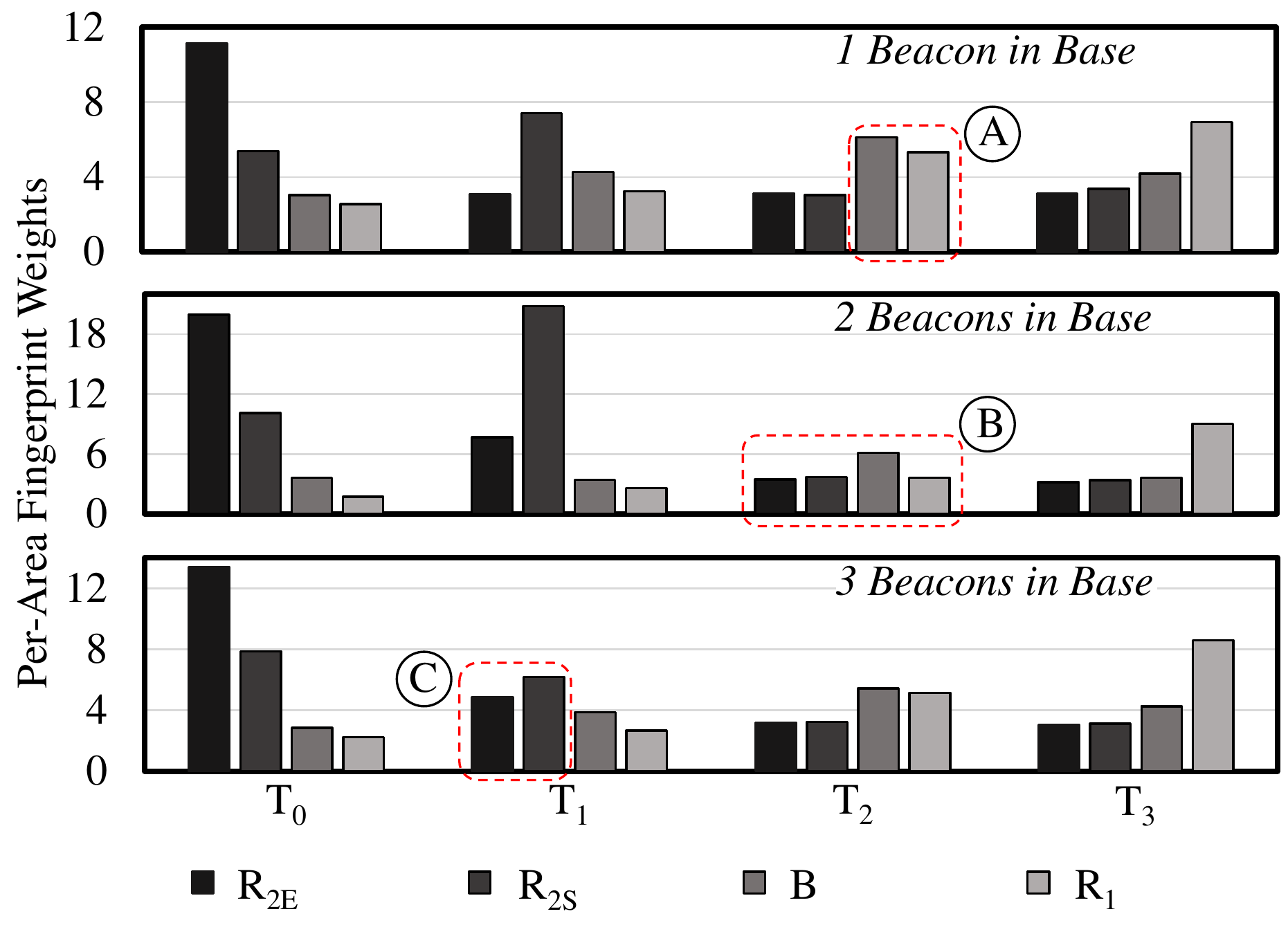}
    \caption{\textbf{Per-area fingerprint weights analysis for 4 different areas with varying number of beacons in Base $B$.}{ \normalfont When the number of beacons in $B$ is 2, we observe the optimal differences in fingerprint weights, thereby obtaining more accurate user localizations.}}
    \label{fig:FW}
\end{figure}

Next, we study the number of beacons required for a given area by deploying varying number of beacons (\{$1$, $2$, $3$\}) in a room on our test-floor ($B$ in Environment~$1$ of floor-plan), to study the effect of their interaction on the fingerprint weights, which determine the location of the user.
Fig.~\ref{fig:FP} illustrates the floor-plan and the beacon deployment used for this experiment, the results of which are illustrated in Fig.~\ref{fig:FW}.
With just $1$ beacon in $B$, although clear localizations can be achieved when the user is in one of the other three rooms, when the user is in $B$, the difference between the fingerprint weights of the highest and the second highest values is quite less, as illustrated by label~\circled{A}.
This distinction becomes clearer when the number of beacons in $B$ is increased to $2$, as shown by~\circled{B}. 
The fingerprint weight difference between the first and second highest values also increase in other rooms, making the localizations more accurate.
When the number of beacons are further increased to $3$, it causes localization problems in other rooms as well, as illustrated by label~\circled{C}.
Therefore, the optimal number of beacons for $B$ is 2, which is the configuration used in all future experiments. 
From this, we can estimate that a beacon is required for every $40m^2-50m^2$ area, to achieve effective localization.
With the timestamps of the data packets and the fingerprint generated for each user, it is possible to successfully localize and track them, with room-level precision, across the environment, given its floor-plan at all time-instances.
As illustrated in Fig.~\ref{fig:FP}, the user device moves to each location, in $15$-second intervals.



Next, we evaluate the multi-user tracking capabilities of our framework by deploying $6$ beacons (colored areas), in a multi-environment scenario as illustrated in Fig.~\ref{fig:LTME}.
We determine the location of the users every $15$-seconds from $T_0$ to $T_3$ and track them through the multi-floor multi-environment scenario by localizing their device at each time instant.
As illustrated, we successfully and simultaneously localize multiple user devices across multiple floors in order to track them.
Furthermore, we have also evaluated the performance overheads associated with the deployed encryption approaches, which have been presented in Table~\ref{tab1}.
As anticipated, the encryption techniques increase the latency of communication between each layer of the network.
However, the required data for the proposed localization algorithm (Algorithm~\ref{Algo1}) is still collected and processed within the time constraint, as discussed in Fig.~\ref{fig:TD}, due to the IoT architecture proposed in the ATLAS framework.


\begin{table}[t]
\centering
\caption{Overview of the performance overhead associated with the deployment of communication protocols with and without encryption.\label{tab1}}
\begin{tabular}{r|cc}
\multicolumn{1}{c|}{\multirow{2}{*}{Communication Layer}} & \multicolumn{2}{c}{Time (in seconds)}            \\ \cline{2-3} 
\multicolumn{1}{c|}{} & \multicolumn{1}{c|}{w/o Encryption} & w/ Encryption \\ \hline
User Localization     & \multicolumn{1}{c|}{1.24}           & 2.11       \\
Database Store        & \multicolumn{1}{c|}{2.13}           & 4.73       \\
First Connection      & \multicolumn{1}{c|}{8.01}           & 10.94     
\end{tabular}
\end{table}
\section{Conclusion and Future Work}
\label{sec:Conc}

In this work, we presented the ATLAS framework, which uses Bluetooth Low Energy beacons in our proposed IoT architecture to anonymously localize and track the users of the smart environment.
Furthermore, to ensure the privacy and the security of its users, we presented an easy-to-adopt secure networking stack, which enables varying layers of encryption across the communication stack.
The ATLAS framework analyzes a combination of signal strength, hashed user IDs, beacon UUIDs, and timestamps to effectively determine the proximal location of each user in the environment, thereby enabling privacy-preserving contact tracing capabilities.
Our framework is open-sourced at \textcolor{blue}{\url{https://atlas-tuw.sourceforge.io}} to enable wide-spread adoption and reproducibility, which can further research and development in this field.

In the next phase of our work, we propose to evaluate the scalability of our ATLAS framework using a larger number of bluetooth beacons and smartphones, followed by the deployment of these beacons across multiple buildings and/other city infrastructures.
This includes the evaluation of bluetooth beacons and smartphones from other manufacturers.


\bibliographystyle{IEEEtran}
\bibliography{References}

\end{document}